\documentclass[lettersize,journal]{IEEEtran}
\usepackage{amsmath,amsfonts}
\usepackage{algorithmic}
\usepackage{array}
\usepackage[caption=false,font=normalsize,labelfont=sf,textfont=sf]{subfig}
\usepackage{textcomp}
\usepackage{stfloats}
\usepackage{url}
\usepackage{verbatim}
\usepackage{graphicx}
\def\BibTeX{{\rm B\kern-.05em{\sc i\kern-.025em b}\kern-.08em
    T\kern-.1667em\lower.7ex\hbox{E}\kern-.125emX}}
\usepackage{balance}
\usepackage{bm} 
\usepackage{booktabs}
\usepackage{multirow}
\usepackage{makecell}
\usepackage{color}

\begin{document}
\title{Enhancing Multimodal Entity and Relation Extraction with 
	Variational Information Bottleneck}

\author{Shiyao Cui, Jiangxia Cao, Xin Cong, Jiawei Sheng, Quangang Li, Tingwen Liu, Jinqiao Shi
\thanks{Manuscript created April, 2023; This work was developed by the IEEE Publication Technology Department. This work is distributed under the \LaTeX \ Project Public License (LPPL) ( http://www.latex-project.org/ ) version 1.3. A copy of the LPPL, version 1.3, is included in the base \LaTeX \ documentation of all distributions of \LaTeX \ released 2003/12/01 or later. The opinions expressed here are entirely that of the author. No warranty is expressed or implied. User assumes all risk. \\ Shiyao Cui, Jiangxia Cao, Xin Cong, Jiawei Sheng, Quangang Li and Tingwen Liu are with the  Institute of Information Engineering, Chinese Academy of Sciences, Beijing 100193, China and the University of Chinese Academy of Sciences, Beijing, China (e-mail: \{cuishiyao,caojiangxia,congxin,shengjiawei,liquangang,liutingwen\}@iie.ac.cn). \\ Jinqiao Shi is with the Beijing University of Posts and Telecommunications, Beijing 100876, China. (e-mail:shijinqiao@bupt.edu.cn) \\ This work has been submitted to the IEEE for possible publication. Copyright may be transferred without notice, after which this version may no longer be accessible.}}

\markboth{Journal of \LaTeX\ Class Files,~Vol.~18, No.~9, September~2020}%
{How to Use the IEEEtran \LaTeX \ Templates}

\maketitle

\begin{abstract}
This paper studies the multimodal named entity recognition (MNER) and multimodal relation extraction (MRE), which are important for multimedia social platform analysis.
The core of MNER and MRE lies in incorporating evident visual information to enhance textual semantics, where two issues inherently demand investigations.
The first issue is modality-noise, where  the task-irrelevant information in each modality may be noises misleading the task prediction.
The second issue is modality-gap, where representations from different modalities are inconsistent, preventing from building the semantic alignment between the text and image.
To address these issues, we propose a novel method for MNER and MRE by \underline{M}ulti\underline{M}odal representation learning  with  \underline{I}nformation \underline{B}ottleneck (MMIB).
For the first issue, a refinement-regularizer probes the information-bottleneck principle to  balance the predictive evidence and noisy information, yielding expressive representations for prediction.
For the second issue, an alignment-regularizer is proposed, where a mutual information-based item works in a contrastive manner to regularize the consistent text-image representations.
To our best knowledge, we are the first to explore variational IB estimation for MNER and MRE.
Experiments show that MMIB achieves the state-of-the-art performances on three public benchmarks.
\end{abstract}


\section{Introduction}

\IEEEPARstart{N}{amed} entity recognition (NER)~\cite{DBLP:conf/acl/YuBP20} and relation extraction (RE)~\cite{DBLP:conf/acl/ZhangYSXLG21} are two major tasks for socia media analysis, benefiting various applications~\cite{DBLP:journals/tkde/GuoZQZXXH22,DBLP:conf/acl/0001WTXXH0Z22}.
Specifically, NER aims to extract entities of interest and RE seeks to decide the semantic relations between entities from unstructured texts, respectively.
Compared with NER and RE in newswire domain, the texts in social media are usually short and accompanied with slangs~\cite{DBLP:conf/mm/WuZCCL020,DBLP:journals/tmm/ZhengWWCL21} , where the inadequate and ambiguous semantics raise the difficulty to content analysis.
Fortunately, with the prevalence of multimodal posts in social media, in recent years,  images in social posts are leveraged to facilitate NER and RE towards social media, namely \textbf{Multimodal NER (MNER)}~\cite{DBLP:conf/acl/JiZCLN18} and \textbf{Multimodal RE (MRE)}~\cite{DBLP:conf/mm/ZhengFFCL021}.

\begin{figure}[t]
	\begin{center}
		\includegraphics[width=8.6cm,height=4.4cm]{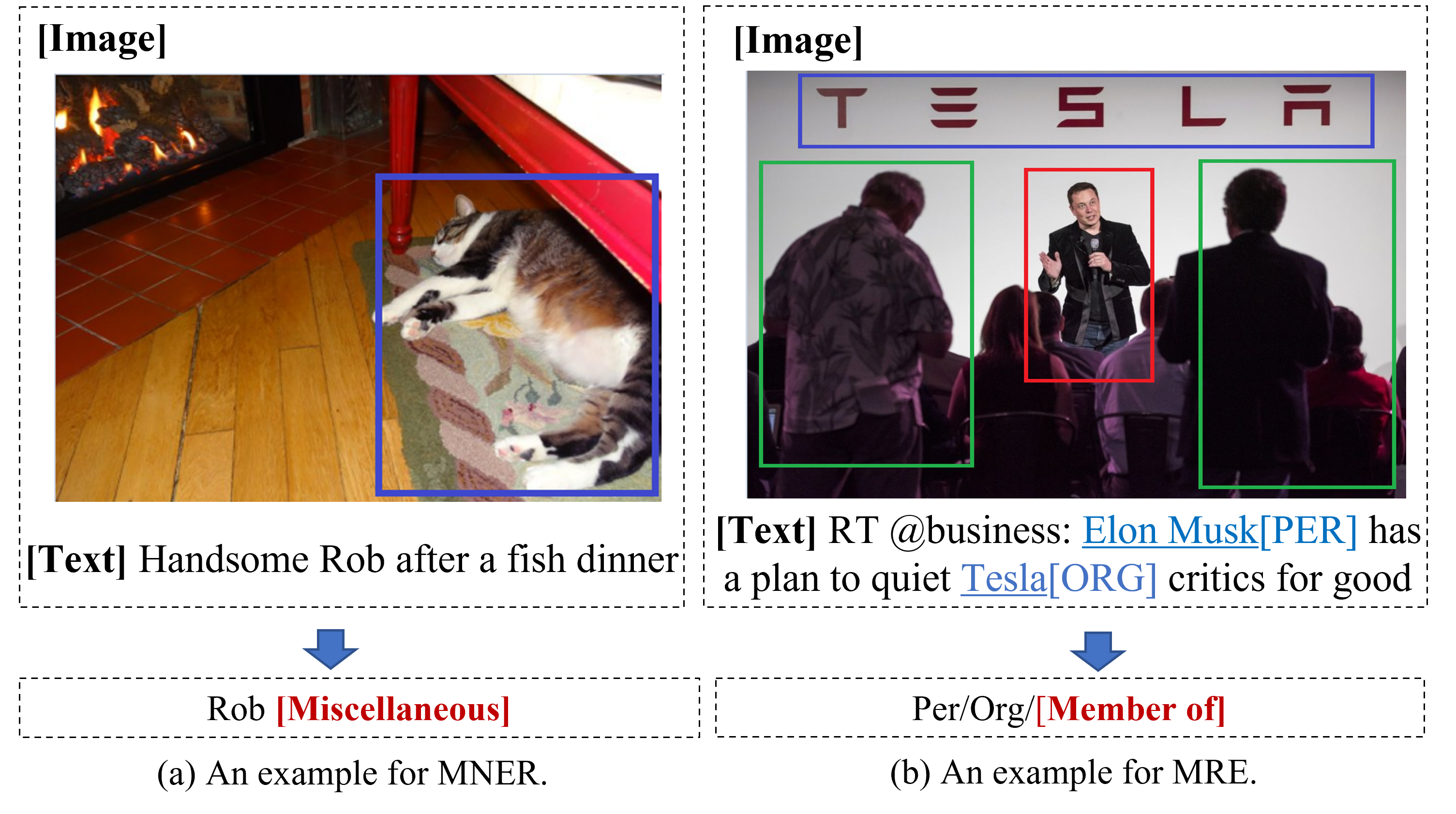}
		\caption{Examples from Twitter dataset for MNER and MRE.}
		\label{fig:example}
	\end{center}
\end{figure}

The core of MNER and MRE lies in incorporating evident visual information from images to enhance the textual semantics in social media.
To achieve this goal, the visual information is integrated with different ways.
Early researchers directly explore the whole image as global-level visual clues.
These methods either encode the raw image as one feature vector~\cite{DBLP:conf/naacl/MoonNC18} for the semantic enhancement, or roughly segment the image into multiple grids for cross-modal interaction~\cite{DBLP:conf/acl/YuBP20,DBLP:conf/wsdm/XuHSW22}.
For the more fine-grained visual clues, the following researchers extract the salient objects in the image and encode them as local-level visual hints~\cite{DBLP:conf/mm/WuZCCL020,DBLP:journals/tmm/ZhengWWCL21,DBLP:conf/mm/ZhengFFCL021,DBLP:conf/aaai/ZhangWLWZZ21,chen-etal-2022-good,DBLP:conf/mm/JiaSSPL00022,DBLP:conf/coling/LuZZZ22,DBLP:journals/corr/abs-2211-14739} to guide the task prediction.
Despite of their remarkable success, two issues are still inadequately addressed.

The first issue is \textbf{modality-noise}, where each modality contains task-irrelevant information, which can not contribute to the final tasks and even be noises misleading the prediction.
Taking the MNER in Figure~\ref{fig:example}(a) as an example.
For the text-modality, the model tends to predict ``Rob'' as an entity of person-type(\texttt{PER}), since the textual noises from some words, like ``handsome'' and ``dinner'', pose characteristics of a person to ``Rob''.
Fortunately, from the overview of the text and image, the ``Cat'' object (e.g. the region in the blue box) in the image is evident to guide the correct prediction.
Nevertheless, for the image-modality, the noises could lie in two-folds: 
1) in the global-level, most regions in the image are not informative to the recognition of the target entity;
2) in the local-level, the corresponding evident region also expresses a more complicated visual semantics (``A cat lies on the carpet'') than we need (just a ``Cat'') exactly.
Here, the redundant information may disturb the model to assigning attention weights for regions in the image, thus impede the final task predictions.
Similarly, in Figure~\ref{fig:example}(b), the redundant noisy regions in green boxes may drive the model to predict one relation between persons, instead of  between an organization and a person.
Hence, it is important to refine information which diminishes the redundant noises and maintains predictive for tasks.

Even though the redundant noises are erased, \textbf{modality-gap} still exists, where the representations of the input text and image are inconsistent.
Specifically, since the textual and visual representations are respectively obtained from different encoders, they maintain different feature spaces and distributions.
Such a modality-gap makes it confusing for the text and image to ``understand'' each other and capture the cross-modality semantic correlations.
For MNER in Figure~\ref{fig:example}(a), ideally, the textual entity ``Rob'' should hold stronger semantic relevance with the ``Cat'' in the blue box than other regions.
However,  the disparity between textual and visual representations prevents from building such alignment, hindering the exploration to predictive visual hints.
So is it for  MRE in Figure~\ref{fig:example}(b),  where the disagreement between semantics expression makes it hard to align the annotated textual entities with visual person/organization objects for relation decision.
Though prior researches~\cite{DBLP:journals/tmm/ZhengWWCL21,wang-etal-2022-ita} convert visual objects into the corresponding textual labels to achieve the consistent semantics expression, it still suffers from the error sensitivity of textual labels produced by external off-the-shelf tools.

In this paper, we propose a novel approach for MNER and MRE by \underline{M}ulti\underline{M}odal Representation learning with variational \underline{I}nformation \underline{B}ottleneck, which is termed as \textbf{MMIB}.
Our method  tackles the issues above by regularizing the data distributions based on the variational auto-encoder~\cite{DBLP:journals/corr/KingmaW13} framework. %
Specifically, for \textbf{modality-noise}, we design a \textit{Refinement-Regularizer (RR)}, which explores the Information-Bottlenack (IB) principle for text-image representation learning.
IB principle aims at deriving representations in terms of \textit{ a  trade-off between having a concise representation with its own information and one with general predictive power}~\cite{DBLP:conf/iclr/AlemiFD017}.
In our task, IB refines text/image representations which are diminished from the noisy information but  predictive for final tasks.
Our devised RR consists of two IB-terms respectively upon the representation learning of the input text and image, yielding robust representations to each modality for prediction.
For \textbf{modality-gap}, an \textit{Alignment-Regularizer (AR)} is proposed.
Since mutual information (MI) could measure the association between two variables, AR drives the consistent text-image representations by maximizing the MI between the representations of the paired text-image  while minimizing that between the unpaired ones.
Based on the regularizers above, we could produce expressive text-image representations to facilitate the final task predictions.

Overall, we summarize our contributions as follows:
\begin{itemize}
	\item We explore a fresh perspective to solve MNER and MRE via representation learning with variational  information bottleneck principle.
	\item We propose a novel method, MMIB, which employs a refinement-regularizer and an alignment-regularizer respectively for modality-noise and modality-gap. %
	\item Experiments show that MMIB achieves new state-of-the-art performances on three MNER and MRE public benchmarks, and extensive analyses verify the effectiveness of our proposed method. 
\end{itemize}

\section{Preliminary}
In this section, we review the preliminary knowledge about the studied tasks and proposed method from three aspects.
\subsection{Task Formulation}
\subsubsection{MNER} Given a sentence  and its associated image, MNER aims to identify the textual named entities and classify the identified entities into the predefined entity types. 
We formulate the task into a sequence labeling paradigm.
Formally, let $(x_1^t, x_2^t, ..., x_n^t)$ denote the input sentence containing $n$ tokens, we aim to predict the corresponding label sequence $ Y = (y_1, y_2, ..., y_n)$ where $y_i$ is a predefined label which follows  BIO-tagging schema~\cite{DBLP:conf/wsdm/XuHSW22,DBLP:conf/eacl/SangV99} so that the entities of interests could be derived.
\subsubsection{MRE} Given a sentence and its associated image, the goal of MRE is to detect the semantic relations between two annotated entities.
Formally, let $(x_1^t, x_2^t, ..., x_n^t)$  denote the sentence containing two entities $E_1=(x_i^t, ..., x_{i+|E_1|-1}^t)$ and $E_2=(x_j^t, ..., x^t_{j+|E_2|-1})$,
the task is formulated as a classification problem to decide the relation types $Y$ between $E_1$ and $E_2$ from the predefined relation types.
\label{mre_explain}
\subsection{Mutual Information}
Before going on, we first introduce Mutual Information (MI) since it is one basic concept for information bottleneck.
MI is a general metric in information theory~\cite{DBLP:journals/bstj/Shannon48}, which measures the strength of association between random quantities, which is defined as follows:

\begin{equation}
	I(\bm{X};\bm{Y}) =  \sum_{y \in \bm{Y}}\sum_{x \in \bm{X}}p(x, y){\rm{log}}\frac{p(x, y)}{p(x)p(y)},
\end{equation}
where $I(\cdot;\cdot)$ denotes the MI between two random variables $\bm{X}$ and $\bm{Y}$, $p(\cdot)$/$p(\cdot, \cdot)$ respectively refer to the marginal/joint \textbf{probability} for samples $x$ and $y$. 
Meanwhile, MI could also be measured by Kullback–Leibler divergence~\cite{DBLP:conf/acl/ZhangZWZCH22,DBLP:conf/icde/CaoSCLW22} between two distributions as:
\begin{equation}
	\label{equ:mi-kl}
	I(\bm{X};\bm{Y}) =  \mathbb{D}_{KL}(\bm{p}(\bm{X}|\bm{Y}) || \bm{p}(\bm{X})),
\end{equation}
where $\bm{p}(\cdot)$/$\bm{p}(\cdot|\cdot)$ denote the prior/posterior \textbf{distribution} for variables $\bm{X}$ and $\bm{Y}$. 

\subsection{Information Bottleneck Principle}

Information Bottleneck (IB), which was proposed by Tishby et al.~\cite{DBLP:conf/itw/TishbyZ15}, aims to derive effective latent features with a tradeoff between having an explicit representation and one with general predictive power~\cite{DBLP:conf/iclr/AlemiFD017}.
Assuming we have the task input information $\bm{X}$ and the task label information $\bm{R}$, the IB is formed as:
\begin{equation}
	\label{eq:ib}
	\mathcal{L}_{IB} = \beta I(\bm{Z};\bm{X}) - I(\bm{Z};\bm{R}),
\end{equation}
Particularly, the goal of Eq.~\ref{eq:ib} is to derive a limited latent variable $\bm{Z}$, which filter redundant information provided by the input $\bm{X}$  while maximally related information to reinvent expected information $\bm{R}$.
The objective function could be explained as follows: 1) \textbf{minimizing} $I(\bm{Z};\bm{X})$ so that $\bm{Z}$ could discard irrelevant parts for task predictions; 2) \textbf
{maximizing} $I(\bm{Z};\bm{R})$ to enforce $\bm{Z}$ to retain the predictive information. 
The $\beta> 0$ is a Lagrangian multiplier to balance the trade-off between the two constraints.
In implementation, since it is intractable to directly optimize the MI-based terms, variational approximation~\cite{DBLP:conf/cogsci/GershmanG14} is widely adopted for optimization to the corresponding objection functions~\cite{DBLP:conf/icde/CaoSCLW22}.

\section{Method}

\begin{figure*}[t]
	\centering
	\includegraphics[width=16cm,height=5.5cm]{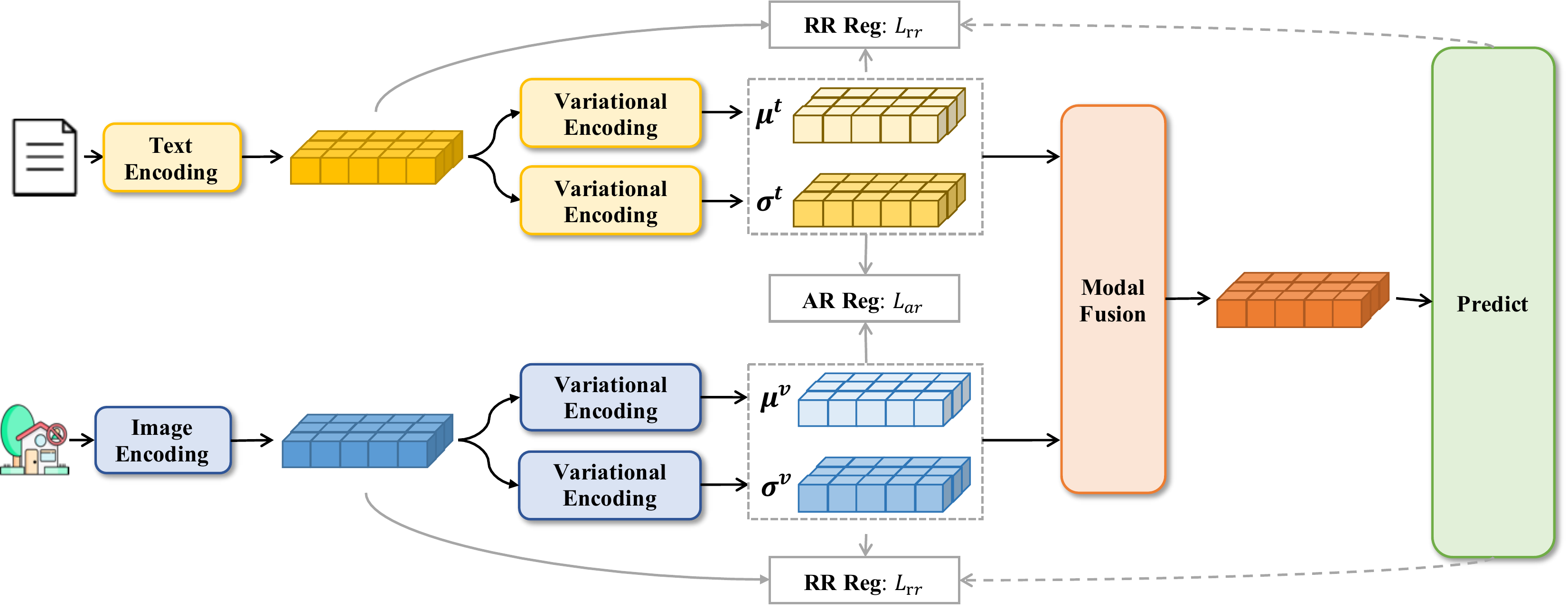}
	\caption{Model Architecture.}
	\label{fig:model-arc}
\end{figure*}

In this section, we introduce MMIB as Figure~\ref{fig:model-arc} shows.
Specifically, MMIB contains three  modules: 1) Encoding module converts the texts and images into the basic real-valued embeddings. 2) Representation learning module   conducts variational encoding with our proposed regularizers. 3) Prediction module conducts the modality fusion and performs the final task predictions.

\subsection{Encoding Module}
\label{sec:encoding}
This module conducts the  encoding towards the input sentence and image, converting them into the distributional representations.

\subsubsection{Text Encoding} To derive the representations for the input sentence, we employ the widely used BERT~\cite{DBLP:conf/naacl/DevlinCLT19} as the textual encoder.
Following  Devlin et al.~\cite{DBLP:conf/naacl/DevlinCLT19}, two special tokens are inserted into each sentence, where $\mathtt{[CLS]}$ and $\mathtt{[SEP]}$ are respectively appended to the beginning and end to the sentence.
Formally, let $X^T = (x_0^t, x_1^t, x_2^t, ..., x_n^t, x_{n+1}^t)$ denote each processed sentence, where $x_0^t$ and $x_{n+1}^t$ respectively denote the inserted $\mathtt{[CLS]}$ and $\mathtt{[SEP]}$.
We feed $X^T$ into BERT, and obtain the output representations as $\bm{X}^T = (\bm{x}_0^t, \bm{x}_1^t, \bm{x}_2^t, ..., \bm{x}_n^t, \bm{x}_{n+1}^t) $, where $\bm{x}_i^t \in \mathbb{R}^d$ is the contextualized representation for the $i_{th}$ token.

\subsubsection{Image Encoding} Given an image $X^V$, it contains visual information from two aspects: (i) the global-level abstract concepts provided by the whole image; (ii) the local-level semantic units provided by the visual objects.
We intend to explore such two-fold visual information, and thus two steps are involved to produce the image representations.
\textbf{Step1. Object Extraction.} 
Following Zhang et al.~\cite{DBLP:conf/aaai/ZhangWLWZZ21}, we employ a visual grounding toolkit~\cite{DBLP:conf/iccv/YangGWHYL19} to extract local objects, which are denoted as $(x^v_1, x^v_2, ..., x^v_m)$. %
\textbf{Step2. Image Representation.} 
Considering the great success of ResNet~\cite{DBLP:conf/cvpr/HeZRS16} in computer vision, we utilize it  as the visual encoder.
Specifically, we first rescale the whole image and its object images into $224 * 224$ pixels, and feed them into ResNet.
Since ResNet produces visual representations in the dimension of $2048$, a linear transformation matrix $\bm{W}_v \in \mathbb{R}^{2048 \times d} $ works to project the image representations into the same dimension as the textual representations.
Finally, we concatenate the image representations together as $\bm{X}^V = (\bm{x}_0^v, \bm{x}_1^v, ..., \bm{x}_m^v)$, where $\bm{x}_0^v \in \mathbb{R}^{1 \times d}$ and $\bm{x}_i^v \in \mathbb{R}^{1 \times d} (i > 0)$ respectively denote the representation of the whole image and $i_{th}$ local object image.

\subsection{Representation Learning Module}
\label{sec:rl}
This module aims to produce the noise-robust and consistent text-image representations  with our proposed  regularizers.
Since these MI-based regularizers are intractable, variational encoding serves for the representation learning.
We will first introduce how the variational encoding works, and then detail the  regularizers.

\subsubsection{Variational Encoding}
\label{sec:vari-encoding}

We devise an encoder in a variational manner for expressive text/image representation learning.
Since the latent text/image representation is compressed  within each modality, the encoder should explore the intra-modal information propagation.
Considering the sufficient ability of Transformer~\cite{DBLP:conf/nips/VaswaniSPUJGKP17} to propagation modeling, we deploy the encoder following the design of a Transformer Layer.
Specifically, \textbf{we name the encoder as ``Attentive Propagation Encoder''}, termed as $ {\rm{APEnc}}(\bm{Q}, \bm{K})$ with a query $\bm{Q}$ and key-value pair $\bm{K}$ as input, where a multi-head attention mechanism is first applied towards two variables as:
\begin{equation}
	\label{equ:apenc-1}
	\begin{aligned}
		& {\rm{MultiHeadAtt}}(\bm{Q}, \bm{K}) = \bm{W}^{'}[{\rm{CA}}_1(\bm{Q}, \bm{K}), ..., {\rm{CA}}_h(\bm{Q}, \bm{K})]^{\rm{T}}, \\
		& {\rm{CA}}_j(\bm{Q}, \bm{K}) = {\rm{Softmax}} \Big(\frac{[\bm{Q} \bm{W}_{qj}  ]  [\bm{K} \bm{W}_{kj}  ]^\mathbf{T} }{\sqrt{d/h}}\Big) [\bm{K}\bm{W}_{vj}  ],
	\end{aligned}
\end{equation}
where ${\rm{CA}}_j$ refers to the $j_{th}$ head of such multi-head attention, $h$ is the number of heads. $\{\bm{W}_{qj},  \bm{W}_{kj}, \bm{W}_{vj}\} \in \mathbb{R}^{d \times (d/h) }$ and $\mathbf{W}^{'} \in \mathbb{R}^{d\times d}$ are all projections parameter matrices.
Then, a fully-connected feed-forward network and a residual layer with layer-normalization are further stacked as follows:  
\begin{equation}
	\label{equ:apenc-2}
	\begin{aligned}
		& \widetilde{\bm{F}} = {{\rm{LayerNorm}}}(\bm{Q} + {\rm{MultiHeadAtt}}(\bm{Q}, \bm{K}) ) \\
		& \bm{F}  = {\rm{LayerNorm}}(\widetilde{\bm{F}} + {\rm{FeedForward}}(\widetilde{\bm{F}})), 
	\end{aligned}
\end{equation}
where $\bm{F}$ is the final output representation of the encoder $ {\rm{APEnc}}(\bm{Q}, \bm{K})$ containing computation from Eq.~\ref{equ:apenc-1} to Eq.~\ref{equ:apenc-2}.

Due to the intractability to MI-based regularizers, the encoder works in a variational manner. 
For simplicity, \textbf{we illustrate the variational encoding procedure  of image representations ${\rm{APEnc}}^V(\bm{X}^V, \bm{X}^V)$ as an example}.
To promote the  information propagation between images, each image $\bm{x}_i^v \in \mathbb{R}^{1 \times d}$ is respectively taken as the query to interact with all images representations $\bm{X}^V \in \mathbb{R}^{(m+1) \times d}$ as key-value pairs.
Specifically, the latent gaussian distributional variable $\bm{z}^v_i$ for each image is produced as follows:
\begin{align}
	\label{equ:image-rep}
	& \bm{\mu}^v_i  = {\rm{APEnc}}^V_{\mu}(\bm{x}_i^v, \bm{X}^V), \\ & (\bm{\sigma}^t_i)^2 = {\rm{exp}} ({\rm{APEnc}}^V_{\sigma} (\bm{x}_i^v, \bm{X}^V)) \\
	& \bm{z}^v_i \sim \mathcal{N}\Big(\bm{\mu}^v_i,  (\bm{\sigma}^v_i)^2 \Big),
\end{align}
where   $\bm{\mu}^v_i \in \mathbb{R}^{1 \times d}$ and $\bm{\sigma}^v_i \in \mathbb{R}^{1 \times d}$ are the mean and variance vector of Gaussian distribution $\mathcal{N}(\bm{\mu}^v_i,  (\bm{\sigma}^v_i)^2 )$, and the representation $\bm{z}^v_i$ of the $i_{th}$ image is sampled from the Gaussian distribution.
However, such a direct sampling process is intractable for back-propagation, thus we adopt the widely used reparameterization  trick~\cite{DBLP:journals/corr/KingmaW13} as a solution.
Mathematically, we first sample $\bm{\epsilon}$ from the normal Gaussian distribution, and then perform the equivalent sampling to derive $\bm{z}^v_i$ as follows:
\begin{equation}
	\label{equ:equ:sample}
	\bm{z}^v_i = \bm{\mu}^v_i +  \bm{\sigma}^v_i \odot \bm{\epsilon},~~~\bm{\epsilon} \sim \mathcal{N}( 0,\text{diag}(\bm{I}))
\end{equation}
where  $\odot$ denotes the element-wise product operation.
Similarly, we could obtain the representations for all images, and correspondingly acquire the visual representations $\bm{Z}^V \in \mathbb{R}^{(m+1) \times d}$ as follows:
\begin{equation}
	\bm{Z}^V = (\bm{z}_0^v, \bm{z}_1^v, ..., \bm{z}_{m}^v).
\end{equation}
\textbf{The above procedure also holds in the case of text encoding via ${\rm{APEnc}}^T(\bm{X}^T, \bm{X}^T)$ and we could obtain}:
\begin{equation}
	\bm{Z}^T = (\bm{z}_0^t, \bm{z}_1^t, ..., \bm{z}_{n + 1}^t).
\end{equation}
%

\subsubsection{Refinement-Regularizer (RR)} 
For the modality-noise, the representations of  text/images should \textbf{be diminished from the task-irrelevant information as much as possible while maximally maintain predictive for final tasks}.
Considering that information-bottleneck (IB) could balance information from two variables, we intend to achieve this goal using IB-based regularizers.
Specifically, we propose a refinement-regularizer $L_{rr}$, which consists of two IB-based terms respectively upon the representation learning of  text $\bm{Z}^T$ and   images $\bm{Z}^V$ as follows:
\begin{equation}
	\label{equ: ibr}
	\small
	\begin{split}
		L_{rr} = & \beta_1 I(\bm{Z}^T;\bm{X}^T) - I(\bm{Z}^{T};\bm{R}) \\ 
		& + \beta_2 I(\bm{Z}^V;\bm{X}^V) - I(\bm{Z}^V;\bm{R}),\\
	\end{split}
\end{equation}
where $\bm{X}^T$, $\bm{X}^V$ respectively denote the original information from input text and image, $\bm{R}$ refers to the ideal task-expected information for predictions.
Considering the independence between $\bm{Z}^{T}$ and $\bm{Z}^{V}$~\cite{DBLP:conf/icde/CaoSCLW22}, the mutual information chain principle could be rewritten as:
\begin{equation}
	\small
	\begin{split}
		I(\bm{Z}^{T};\bm{R}) + I(\bm{Z}^V;\bm{R})  = & I(\bm{Z}^{T};\bm{R} | \bm{Z}^V) + I(\bm{Z}^V;\bm{R}) \\
		= &I(\bm{Z}^{T},\bm{Z}^V;\bm{R}).\\
	\end{split}
	\label{chain}
\end{equation}
Accordingly, $L_{rr}$ in Eq.\ref{equ: ibr} could be simplified as follows:
\begin{equation}
	\label{equ:rr-reg}
	\small
	\begin{split}
		L_{rr} =\underbrace{ \beta_1 I(\bm{Z}^{T};\bm{X}^T) + \beta_2 I(\bm{Z}^V;\bm{X}^V)}_{\rm{Minimality}} - \underbrace{I(\bm{Z}^{T},\bm{Z}^V;\bm{R})}_{\rm{Reconstruction}},\\
	\end{split}
\end{equation}
where $L_{rr}$ could be optimized by (i) minimizing the former two terms so that the  text/image representations are \textbf{compressed from their original modality apart from noises}; (ii) maximizing the reconstruction term  to \textbf{encourage the predictive representations towards task predictions}.
More specifically, $L_{rr}$ refines the text/image representations to maintain evident  but limits the disturbing noises.
\textbf{For tractable objective function of $L_{rr}$ , please refer to Section~\ref{sec:rr-function}.}

\subsubsection{Alignment-Regularizer (AR)}
For the modality-gap problem, the representations of  text/images should be consistent with each other.
We  achieve this goal by \textbf{enhancing the mutual information (MI) between $\bm{Z}^{T}$ and $\bm{Z}^{V}$ for each input text-image pair}.
In this way, the text-image representations are encouraged to maximumly agree with each other, since MI  could inherently measure the association between two variables.
Specifically, we devise a MI-based alignment-regularizer (AR), $L_{ar}$, as follows:
\begin{equation}
	\label{equ:ar}
	L_{ar} = - \underbrace{I(\bm{Z}^{T}; \bm{Z}^V)}_{\rm{Maximality}},
\end{equation}
where $L_{ar}$ is optimized by maximizing  $I(\bm{Z}^{T}; \bm{Z}^V)$ for the agreement between representations of paired text/image.
\textbf{For the tractable objective function of Eq.~\ref{equ:ar}, please refer to Section~\ref{sec:ar-function}.}

\subsection{Prediction Module}
In this module, we first conduct modality fusion to incorporate information from both modalities, and then respectively perform specific task predictions.

\subsubsection{Modality Fusion}
To derive the image-aware text representations for the final task predictions, the information interaction between modalities is probed. 
\textbf{The above mentioned attentive propagation encoder (${\rm{APEnc}}$) is again leveraged here, but the key difference lies in that the query and key-value pairs are taken  from different modalities}.
To exploit the bi-directional interactions between modalities, ${\rm{APEnc}}$ works in a coupled way, where two encoders respectively take the image and text as query.
Specifically, for $\bm{Z}^V $ and $\bm{Z}^T$, the first ${\rm{APEnc}}^{{\rm{V2T}}}$ aggregates information from texts to derive the text-attended image representations $\bm{B} \in \mathbb{R}^{(m+1) \times d}$  as follows:
\begin{equation}
	\bm{B} = {\rm{APEnc}}^{{\rm{V2T}}}(\bm{Z}^V, \bm{Z}^T).
\end{equation}
Then, to enhance the textual representations with visual information, the second ${\rm{APEnc}}^{{\rm{T2V}}}$ serves to derive the image-aware text representations as follows:
\begin{equation}
	\bm{C} = {\rm{APEnc}}^{{\rm{T2V}}}(\bm{Z}^T, \bm{B}),
\end{equation}
where $\bm{C}=(\bm{c}_0, \bm{c}_1, ..., \bm{c}_n, \bm{c}_{n+1}) $ is the image-aware text representations for the final task predictions.

Since we formulate MNER as a text-level sequence labeling problem, we utilize a standard Conditional Random Field (CRF) to derive the a corresponding label sequence $Y = (y_0, y_1, y_2, ..., y_n, y_{n+1})$ to recognize entities of interest.
Given the final hidden representation of input the sentence, we feed it into CRF,  deriving the  probability to a label sequence  $Y$ as:
\begin{equation}
	\label{equ:crf}
	p(Y|X^T) = \frac{{\rm{exp}}(\sum\limits_{i=0}^{n+1}( \bm{w}^{crf}_{y_i} \bm{c}_{i} + {\rm{Trans}}{(y_{i-1}, y_i)}))}  {\sum\limits_{Y^{'} \in \bm{Y}} {\rm{exp}}(\sum\limits_{i=0}^{n+1}(\bm{w}^{crf}_{y_i} \bm{c}_{i}^{'} + {\rm{Trans}}{(y_{i-1}^{'}, y_i^{'})})))}   ,
\end{equation}
where $\bm{w}^{crf}_{y_i}$ is a parameter vector computing the emission score from token $\bm{c}_{i}$ to label $y_i$,  ${\rm{Trans}}{(y_{i-1}, y_i)}$ is a learnable transition function from $y_{i-1}$ to $y_i$, $\bm{Y}$ is the all possible label sequences set. 
\subsubsection{MRE Prediction} MRE is formulated as a classification problem. 
We derive each annotated entity representation with pooling upon its contained tokens from $\bm{C}$, and further denote the representations of given entities $E_1$ and $E_2$  (explained in \S~\ref{mre_explain})  as $\bm{E}_{1}$ and $\bm{E}_{2}$.
Assuming that the gold label $Y$ is the $k_{th}$ relation  among all relation types, its corresponding probability  is:
\begin{equation}
	\label{equ:re-cls}
	p(Y|X^T, E_{1}, E_{2}) = {\rm{Softmax}}(\bm{W}([\bm{E}_{1} \oplus \bm{E}_{2}]) + \bm{b})[k],
\end{equation}
where $\oplus$ is the concatenation, $\bm{W}$ and $\bm{b}$ are learnable parameters.

\subsection{Training Objective}

The training objective to the MMIB consists of three components: 1) Objective function to the refinement-regularizer; 2) Objective function to the alignment-regularizer; 3) Objective function to the task predictions. 
In the following, we will respectively detail them.

\subsubsection{Refinement-Regularizer Objective Function}
\label{sec:rr-function}
The optimization to $L_{rr}$ in Eq.~\ref{equ:rr-reg} could be performed by minimizing the MI-based two terms and maximizing the reconstruction term as follows.

\textbf{Minimization to the MI-based minimality terms.} Considering the challenge of exact computation to MI~\cite{DBLP:conf/acl/ColomboPC20}, we estimate these two terms based on a \textbf{variational upper bound} through Kullback-Leibler.
Since the two terms, $I(\bm{Z}^{T};\bm{X}^T), I(\bm{Z}^V;\bm{X}^V)$, share the similar forms, we take $I(\bm{Z}^V;\bm{X}^V)$ as an example to illustrate the process. 
Specifically, following Eq.~\ref{equ:mi-kl}, we first measure $I(\bm{Z}^V;\bm{X}^V)$ as follows: 
\begin{equation}
	\label{equ:mi-kl2}
	\centering
	I(\bm{Z}^{V};\bm{X}^V) = \mathbb{D}_{KL}(\bm{p}_{\theta}^V(\bm{Z}^{V}|\bm{X}^V) || \bm{p}(\bm{Z}^{V})).
\end{equation}
Then, following prior variational works~\cite{DBLP:conf/icde/CaoSCLW22,DBLP:conf/iclr/HigginsMPBGBML17,DBLP:conf/nips/ChenLGD18}, the prior distribution $\bm{p}(\bm{Z}^{V})$ could be estimated as  normal Gaussian distribution $\mathcal{N}\big(0, \text{diag}(\bm{I})\big)$.
For the the posterior distribution $\bm{p}_{\theta}^V(\bm{Z}^{V}|\bm{X}^V)$, it could be approximated by a variational posterior distribution $\bm{q}_{\phi}^V(\bm{Z}^{V}|\bm{X}^V)$.
Accordingly, we could obtain the upper bound of $I(\bm{Z}^{V};\bm{X}^V)$ as:
\begin{equation}
	\label{equ:ib-obj}
	\centering
	\small
	\begin{aligned}
		I(\bm{Z}^{V};\bm{X}^V) & \leq \mathbb{D}_{KL}(\bm{q}_{\phi}^V(\bm{Z}^{V}|\bm{X}^V) || \bm{p}(\bm{Z}^{V})) \\
		& = \sum_{i=0}^{m+1}\mathbb{D}_{KL}\Big(\mathcal{N}\big(\bm{\mu}^{v}_i, [\text{diag}(\bm{\sigma}^{v}_i)]^2\big) || \mathcal{N}\big(0, \text{diag}(\bm{I})\big)\Big) ,
	\end{aligned}
\end{equation}
where $\bm{q}_{\phi}^v$ refers to the parameters of variational encoding in Eq.~\ref{equ:image-rep}. $I(\bm{Z}^{V};\bm{X}^V)$ is optimized by by minimizing its upper bound in Eq.~\ref{equ:ib-obj} as the objective function.
Similarly, the object function of $I(\bm{Z}^{T};\bm{X}^T)$ could also be obtained by following the process from Eq.~\ref{equ:mi-kl2} to Eq.~\ref{equ:ib-obj}.

\textbf{Maximization to the reconstruction term.} The goal of the reconstruction term  is to develop the ideal task-expected information $\bm{R}$ using the latent variables.
We measure the quality of the reconstructed information via the probability to how the learned text/image representations could predict the true label  as follows:
\begin{equation}
	\centering
	\small
	I(\bm{Z}^{T},\bm{Z}^V;\bm{R}) \cong \mathbb{E}_{\bm{p}_\theta^T(\mathbf{Z}^T|\bm{X}^T)\bm{p}_\theta^V(\bm{Z}^{V}|\bm{X}^V)}[\log{p(Y|\bm{Z}^{T},\bm{Z}^V)}],
\end{equation}
where $Y$ refers to the targets (i.e. the gold label sequence for MNER and gold relation for MRE). $\bm{p}_\theta^T(\bm{Z}^T|\bm{X}^T), \bm{p}_\theta^V(\bm{Z}^{V}|\bm{X}^V)$ are estimated via variational encoding $\bm{q}_\phi^T(\bm{Z}^T|\bm{X}^T)\bm{q}_\phi^V(\bm{Z}^{V}|\mathbf{X}^V)$ and we could obtain \textbf{the variational lower bound} of $I(\bm{Z}^{T},\bm{Z}^V;\bm{R})$ as follows:
\begin{equation}
\label{equ:recons}
\footnotesize
\begin{aligned}
	I(\bm{Z}^{T},\bm{Z}^V;\bm{R}) & \geq \mathbb{E}_{\bm{q}_\phi^T(\bm{Z}^T|\bm{X}^T)\bm{q}_\phi^V(\bm{Z}^{V}|\mathbf{X}^V)}[\log{p(Y|\bm{Z}^{T},\bm{Z}^V)}],\\
	&={\rm{log}}(p(Y|\bm{Z}^{T},\bm{Z}^V)) + \!\!\!\!\!\!\!\!\sum_{Y'\in \bm{Y},Y'\neq Y}\!\!\!\!\!\!{\rm{log}}(1 - p( Y'|\bm{Z}^{T},\bm{Z}^V))
\end{aligned}
\end{equation}
where $\bm{Y}$ refers to the label space of task predictions,  $Y$ refers to the true label while $Y'$ as the false labels.
To this end, we employ $p(\cdot | \cdot)$ in Eq.~\ref{equ:crf} and Eq.~\ref{equ:re-cls} respectively as $p(Y|\bm{Z}^{T},\bm{Z}^V)$  for MNER and MRE.
Hence, the reconstruction term could be optimized using the corresponding \textbf{task objective function in Section~\ref{sec:task-objective-function}}.

\subsubsection{Alignment-Regularizer Objective Function}
\label{sec:ar-function}
The core of optimization to $L_{ar}$  lies in the \textbf{maximization to the MI-based maximality term} in Eq.~\ref{equ:ar}.
We optimize $L_{ar}$ following infomax~\cite{DBLP:conf/iclr/VelickovicFHLBH19}, where a neural networks measure MI in a contrastive manner. 
Correspondingly, we build a discriminator $\mathcal{D}$ to measure the degree of consistence between text-image representations as:
\begin{equation}
I(\bm{Z}^{T}; \bm{Z}^V) = \mathbb{E}_{\bm{p}_\theta^T(\bm{Z}^T|\bm{X}^T)\bm{p}_\theta^V(\bm{Z}^{V}|\bm{X}^V)}[\log{\mathcal{D}(\bm{Z}^{T}; \bm{Z}^V)}].
\end{equation}
Since the direct optimization to MI-based $L_{ar}$ is intractable, we utilize its \textbf{lower bound estimated via variational encoding} as the objective function as follows:
\begin{equation}
\small
\begin{aligned}
I(\bm{Z}^{T}; \bm{Z}^V) & \geq \mathbb{E}_{\bm{q}_\phi^T(\bm{Z}^T|\bm{X}^T)\bm{q}_\phi^V(\bm{Z}^{V}|\bm{X}^V)}[\log{\mathcal{D}(\bm{Z}^{T}; \bm{Z}^V)}] \\
& = \!\log \big(\mathcal{D}(\bm{Z}^{T}, \bm{Z}^{V+})\big) \! + \! \log\big(1 - \mathcal{D}(\bm{Z}^{T}, \bm{Z}^{V-})\big),
\end{aligned}
\end{equation}
where  $\bm{q}_\phi^T$ / $\bm{q}_\phi^V$ denote the parameters of the variational encoding in Section~\ref{sec:vari-encoding},  $(\bm{Z}^{T}, \bm{Z}^{V+})$ refers to the sampled text-image representations belonging to the same input pair, otherwise the in-batch negatives $(\bm{Z}^{T}, \bm{Z}^{V-})$ in contrastive mutual information.
Since the shape of $\bm{Z}^{T} \in \mathbb{R}^{(n+2) \times d}$ and $\bm{Z}^{V} \in \mathbb{R}^{(m+1) \times d}$ does not match, we conduct mean-pooling to them for computation, namely: 
\begin{equation}
\centering
\small
\label{equ:disc}
\mathcal{D}(\bm{Z}^{T}, \bm{Z}^V) = {\rm{Sigmoid}}({\rm{MLP}}({\rm{Pooling}}( \bm{Z}^{T}) \oplus {\rm{Pooling}}(\bm{Z}^V ))),
\end{equation}
where ${\rm{Pooling}}( \bm{Z}^{T}) \in \mathbb{R}^{d}$ , ${\rm{Pooling}}( \bm{Z}^{V}) \in \mathbb{R}^{d}$ and $\oplus$ denotes the concatenation operation between vectors. The tractable function could be optimized using a standard binary cross-entropy.

\subsubsection{Task Objective Function}
\label{sec:task-objective-function}

We respectively introduce the task-specific objection function for MNER and MRE.

\textbf{MNER.} For MNER, given the sentence $X^T$ and its golden sequence labels $Y$, we could obtain the probability of $Y$ as $p(Y|X^T)$ from Eq.~\ref{equ:crf} and compute the corresponding loss $L_{ner}$ as follows:
\begin{equation}
\label{equ:ner-loss}
L^{ner}_{task} = -{\rm{log}}(p(Y|X^T)).
\end{equation}

\textbf{MRE.} For MRE, given the sentence $X^T$ and the annotated head/tail entity $E_{1} / E_{2}$, we could obtain the probability $p(Y|X^T, E_{1}, E_{2})$ from Eq.~\ref{equ:re-cls} and the objection function could be derived as follows:
\begin{equation}
\label{equ:re-loss}
L^{re}_{task} = -{\rm{log}}(p(Y|X^T, E_{1}, E_{2})).
\end{equation}

\subsubsection{Overall Objective Function}
Based on the functions above, we derive the final objection function $L$ for  MMIB as follows:
\begin{equation}
L = L_{rr} + L_{ar} + L_{task},
\end{equation}
where the all the minimality terms and reconstruction term in $L_{rr}$ are tractably optimized by Eq.~\ref{equ:ib-obj} and Eq.~\ref{equ:recons}, the maximality term in $L_{ar}$ is tractably optimized by Eq.~\ref{equ:disc}.
$L_{task}$ is respectively optimized by Eq.~\ref{equ:ner-loss} and Eq.~\ref{equ:re-loss} for MNER and MRE.

\section{Experiments}

\subsection{Experimental Setup}
\subsubsection{Dataset} We conduct experiments on \textit{Twitter-2015}~\cite{DBLP:conf/aaai/0001FLH18} and \textit{Twitter-2017}~\cite{DBLP:conf/acl/JiZCLN18} for MNER, and \textit{MNRE}~\cite{DBLP:conf/mm/ZhengFFCL021} dataset for MRE.
These datasets are collected from  multimodal posts on Twitter, and each twitter post consists of one piece of text and a image.
\textit{Twitter-2015} and \textit{Twitter-2017} both contain four types of entities: Person (\textbf{PER}), Location (\textbf{LOC}), Organization (\textbf{ORG}) and Miscellaneous (\textbf{MISC}).
Table~\ref{tab:ner-data} shows the number of entities for each type and the division of multimodal tweets in the training, development, and test sets of the two dataset.
The \textit{MNRE} dataset contains 9,201 sentence-image pairs with 23 relation categories.
Table~\ref{tab:re-data} shows the detailed statistics and we compare it with the widely used relation extraction dataset SemEval-2010 Task 8~\cite{DBLP:journals/corr/abs-1911-10422}, reflecting the effectiveness of MNRE for task evaluation.

\subsubsection{Metric} We adopt the same evaluation metric as our baselines. For \textbf{MNER}, an entity is correctly recognized when its span and entity type both match the gold answer. For \textbf{MRE}, the relation between a pair of entities is correctly extracted when the predicted relation type meets the gold answer. We utilize the evaluation code released by Chen et al.~\cite{chen-etal-2022-good}, where the Precision (\textbf{P}), Recall (\textbf{R}) and F1-score (\textbf{F1}) are used for performance evaluation.

\begin{table}[t]
	\footnotesize
	\centering
	\caption{ The Basic Statistics of Twitter-2015 and Twitter-2017 }
	\setlength\tabcolsep{4.7pt}{
		\begin{tabular}{c|ccc|ccc}
			\toprule
			\multirow{2}{*}{Entity Type} & \multicolumn{3}{c|}{Twitter15} & \multicolumn{3}{c}{Twitter17}  \\
			\multicolumn{1}{c|}{} &  \bf Train & \bf Dev & \bf Test &  \bf Train & \bf Dev & \bf Test  \\
			\midrule
			PER & 2217 & 552 & 1816 &  2943 & 626 & 621 \\
			LOC &  2091 & 522 & 1697 & 731 & 173 & 178\\
			ORG & 928 & 247  & 939 &   1674 & 375 & 395 \\
			MISC & 940 & 225 & 726 & 701 & 150 & 157 \\
			\midrule
			Total & 7176 & 1546 & 5078 & 6049 & 1324 & 1351 \\
			\midrule
			Tweets & 4000 & 1000 & 3257 & 3373 & 723 & 723 \\
			\bottomrule
		\end{tabular}
	}
	\label{tab:ner-data}
\end{table}

\begin{table}[t]
	\footnotesize
	\centering
	\caption{ The Statistics of MNRE Compared to SemEval-2010 Dataset }
	\setlength\tabcolsep{5.0pt}{
		\begin{tabular}{ccccc}
			\toprule
			Dataset & Sentence & Entity & Relation & Image  \\
			\midrule
			 MNRE & 9,201 & 30,970 & 23 &  9,201 \\
			SemEval-2010 & 10,717 &  21,434 & 9 & 0 \\
			\bottomrule
		\end{tabular}
	}
	\label{tab:re-data}
\end{table}

\subsubsection{Hyperparameters} We adopt BERT$_{base}$ and ResNet50 as the textual and visual encoders for a fair comparision with previous study~\cite{chen-etal-2022-good}.
Accordingly, the hidden size of the all latent variables and representations are 768.
Parameter optimization are performed using AdamW~\cite{Loshchilov2019DecoupledWD}, where the decay is 0.01, the  learning rate is 3e-5 and the batch size is 8.
We manually tune the Lagrangian multiplier $\beta_1$,$\beta_2$ and achieve the best results with $\beta_1=0.1$ and $\beta_2=0.1$.
The maximum length of the input sentence is  128 for MNER and 80 for MRE by cutting the longer ones and padding the shorter ones.
The model implementation code has been packaged and uploaded as supplementary materials for reproductivity check.

\begin{table}[t]
	\footnotesize
	\centering
	\caption{ Overall MNER results. }
	\setlength\tabcolsep{3.2pt}{
		\begin{tabular}{c|ccc|ccc}
			\toprule
			\multirow{2}{*}{\bf Method} & \multicolumn{3}{c|}{Twitter15} & \multicolumn{3}{c}{Twitter17}  \\
			\multicolumn{1}{c|}{} &  \bf P & \bf R & \bf F1 &  \bf P & \bf R & \bf F1   \\
			\midrule
			CNN-BiLSTM-CRF & 66.24 & 68.09 & 67.15 &  80.00 & 78.76 & 79.37 \\
			HBiLSTM-CRF &  70.32 & 68.05 & 69.17 & 82.69 & 78.16 & 80.37 \\
			
			T-NER & 69.54 & 68.65  & 69.09 &   - & - & -  \\
			\midrule
			GVATT & 73.96 & 67.90 & 70.80 & 83.41 & 80.38 & 81.87 \\
			AdapCoAtt &  72.75 & 68.74 & 70.69 & 84.16 & 80.24 & 82.15 \\
			UMT  &  71.67 & 75.23 & 73.41 & 85.28 & 85.34 & 85.31  \\ 
			FMIT & 74.18 & 75.03 &74.60 & 85.55 & 85.29 & 85.42 \\
			\midrule
			UMGF & \bf 74.49 & 75.21 & 74.85  & 86.54 & 84.50 & 85.51 \\ 
			MAF & 71.86 & 75.10 & 73.42 & 86.13 & 86.38 &  86.25  \\ 
			\midrule
			HVPNeT & 73.87 & 76.82 & 75.32 & 85.84 & 87.93 & 86.87 \\
			\midrule
			\bf MMIB &   74.44 & \bf 77.68 & \bf 76.02 & \bf 87.34 &  87.86 &  \bf 87.60\\
			\bottomrule
		\end{tabular}
	}
	\label{tab:ner-performance}
\end{table}

\subsection{Baselines}
For a comprehensive comparison, we compare our method with four groups of baselines.

\textbf{Text-based baselines}:   To verify the effectiveness of introducing visual images, we choose the baselines which conduct the both tasks without incorporating the visual information. The \textbf{NER} baselines contain 1) \textbf{CNNBiLSTM-CRF}~\cite{DBLP:conf/acl/MaH16} utilizes  word- and character-level representations via BiLSTM and CNN for NER;  
2) \textbf{HBiLSTMCRF}~\cite{DBLP:conf/naacl/LampleBSKD16} replaces the CNN in CNNBiLSTM-CRF with an LSTM; 
3) \textbf{T-NER}~\cite{DBLP:conf/emnlp/RitterCME11} is a Twitter-specific NER system with various features to boost performance, and we refer to its results on our used datasets from Xu et al.~\cite{DBLP:conf/wsdm/XuHSW22}. The \textbf{RE} baselines involve 1) \textbf{PCNN}~\cite{DBLP:conf/emnlp/ZengLC015} uses convolutional networks with piecewise pooling; 2) \textbf{MTB}~\cite{DBLP:conf/acl/SoaresFLK19} is a  RE-oriented pre-training model based on BERT.
\\
\textbf{Vanilla multimodal baselines}: To confirm the necessity of solving modality-noise and modality-gap, we choose vanilla multimodal baselines which ignore these two issues.
Specifically, the \textbf{MNER} baselines contain 1) \textbf{GVATT}~\cite{DBLP:conf/acl/JiZCLN18} utilizes attention mechanism to combine the image- and text-level information;  2) \textbf{AdapCoAtt}~\cite{DBLP:conf/aaai/0001FLH18} designs an adaptive co-attention network to explore visual information; 
3) \textbf{UMT}~\cite{DBLP:conf/acl/YuJYX20} proposes a multimodal interaction module to obtain image-aware word representations;
4) \textbf{FMIT}~\cite{DBLP:conf/coling/LuZZZ22} designs a flat multi-modal interaction transformer (FMIT) layer where a unified lattice structure is used for cross-modality interaction. We cite the performance of FMIT with one FMIT layer for a fair comparison with MMIB.
For \textbf{MRE}, it contains 1)  \textbf{VisualBERT}~\cite{DBLP:journals/corr/abs-1908-03557}, where Chen et al.~\cite{chen-etal-2022-good} utilize the pre-trained multimodal model for text-image encoding and fusion. 2) \textbf{BERT-SG}~\cite{DBLP:conf/mm/ZhengFFCL021}  directly incorporates visual features extracted using a Scene Graph (SG) Tool~\cite{DBLP:conf/cvpr/TangNHSZ20}.
\\
\textbf{Multimodal baselines considering modality-gap.} We choose baselines which tend to derive the consistent text-image representations. For \textbf{MNER}, it contains
1) \textbf{UMGF}~\cite{DBLP:conf/aaai/ZhangWLWZZ21} builds a multimodal graph for semantic alignment.
2) \textbf{MAF}~\cite{DBLP:conf/wsdm/XuHSW22} employs contrastive learning method for  consistent representations.
For \textbf{MRE}, \textbf{MEGA}~\cite{DBLP:conf/mm/ZhengFFCL021} develops a dual graph for semantic agreement.
\\
\textbf{Multimodal baselines considering modality-noise.} To validate the superiority of MMIB for tackling modality-noise, we choose baselines which attempt to allevite the issue. For \textbf{MNER},  \textbf{HVPNet}~\cite{chen-etal-2022-good} alleviates the noises of irrelevant visual objects by exploring the hierarchical visual features as pluggable visual prefix.
For \textbf{MRE}, \textbf{MKGformer}~\cite{DBLP:conf/sigir/ChenZLDTXHSC22} utilizes a correlation-aware fusion module to alleviate the noisy information.

Note that the original results of UMT, UMGF and MEGA only involve one task, and we refer to their results from Chen et al.~\cite{chen-etal-2022-good}. 

\subsection{Main Results}

Table~\ref{tab:ner-performance} and Table~\ref{tab:re-performance} respectively show the final model performances upon MNER and MRE, with significant difference (p$<$0.05) between MMIB and all baselines. Reading from the resutls, we have observations as follows.

1) \textbf{Visual information indeed helps to boost performances.} As multimodal methods generally outperform the text-based ones upon two tasks, we could see the necessarity of incorporating visual information for semantics enhancement. 
However, the performance improvements are still limitted, which demands further exploration to multimodal methods.

2) \textbf{Alleviating the modality-noise and bridging the modality-gap could bring performance gain to multimodal methods}. In these two tables, the latter two groups of baselines perform better than the vanilla multimodal ones. This demonstrates the importance of consistent image-text representations and noise-robust modality information.
Despite of this, the final performances upon two tasks are still barely satisfactory, which calls for better strategies for tackling these two issues.

3) \textbf{Our proposed MMIB achieves the best results upon two tasks.} 
As the two tables show, MMIB outperforms the prior multimodal baselines considering the modality-gap and modality-noise problems.
This demonstrates the effectiveness of our method to handle the two issues.
We attribute such performance advantage to that the refinement-regularizer and alignment-regularizer could effectively derive the ``good'' text-image representations for the final task predictions.

\begin{table}[t]
	\centering
	\caption{Overall MRE performances. }
	\setlength\tabcolsep{15.2pt}{
		\begin{tabular}{c|ccc}
			\toprule
			\multirow{2}{*}{\bf Method} & \multicolumn{3}{c}{MRE} \\
			&  \bf P & \bf R & \bf F1   \\
			\midrule
			PCNN & 62.85 & 49.69 & 55.49 \\
			MTB &   64.46 & 57.81 & 60.86  \\
			\midrule
			AdapCoAtt & 64.67 & 57.98 & 61.14 \\
			VisualBERT 	& 57.15 & 59.48  & 58.30 \\
			BERT+SG  & 62.95 & 62.65 & 62.80 	\\
			UMT & 62.93 & 63.88 & 63.46\\
			\midrule
			UMGF   & 64.38 & 66.23 & 65.29\\	
			MEGA  & 64.51 & 68.44 & 66.41  \\
			\midrule
			HVPNeT &  \bf 83.64  & 80.78 & 81.85 \\
			MKGformer & 82.67 & 81.25 & 81.95 \\
			\midrule
			\textbf{MMIB} & 83.49 & \bf 82.97 & \bf 83.23 \\
			\bottomrule
		\end{tabular}
	}
	\label{tab:re-performance}
\end{table}

\section{Analysis and Discussion}

\subsection{Ablation Study}

To probe how each proposed component contributes to the final performances, we conduct ablation study upon two tasks in three benchmarks. 
Following prior studies~\cite{DBLP:conf/wsdm/XuHSW22,DBLP:conf/aaai/ZhangWLWZZ21,chen-etal-2022-good}, we report ablation performances in the test set in Table~\ref{tab:test-ablation}. 
Specifically, we respectively remove the proposed regularizers to confirm their effectiveness, and present the corresponding analysis as follows.

1) -- Refinement-regularizer ($L_{rr}$): Removing the refinement-regularizers  degrades the final performances upon three benchmarks, revealing the importance of limiting task-relevant noisy information.
Without $L_{rr}$, the noisy information in both modalities may disturb the incorporation to evident visual information, thus exert negative impacts to the final performances.

2) -- Alignment-regularizer ($L_{ar}$): When the alignment-regularizer is removed, the model performances hurt upon all benchmarks.
This phenomenon indicates the necessarity of bridging the modality-gap and verifies the effectiveness of $L_{ar}$ to drive the consistent text-image representations. 

3) -- $L_{rr}$ \& $L_{ar}$: The final performances severely drop when we simultaneously remove all the regularizers. This reveals that these regularizers are both functional and  collaboratively work with each other to boost the final performances.

\begin{table}[t]
	\centering
	\caption{Ablation study.}
	\setlength\tabcolsep{6.3pt}{
		\begin{tabular}{lccccc}
			\toprule
			\multirow{2}{*}{Ablation} & \multirow{2}{*}{Dataset} & \multicolumn{4}{c}{Metrics}   \\ 
			\cline{3-6} 
			&& $\mathbf{P}$ & $\mathbf{R}$ & $\mathbf{F1}$ & $\Delta$ $\mathbf{F1}$  \\
			\midrule
			\multirow{3}{*}{MMIB} & Twitter15  & 74.44 & 77.68 & 76.02 & -\\
			\multirow{3}{*}{} & Twitter17 & 87.34  &  87.86   &  87.60 & -\\
			\multirow{3}{*}{} & MRE & 83.49 & 82.97 & 83.23 & -\\
			\midrule
			\multirow{3}{*}{-- $L_{rr}$}  & Twitter15   &  73.36  & 76.61 & 74.95 & -1.07\\
			\multirow{3}{*}{}  & Twitter17 & 86.67  & 87.56  & 87.11 & -0.49 \\
			\multirow{3}{*}{} & MRE & 82.10 & 81.71 & 81.91 & -1.32 \\
			\midrule
			\multirow{3}{*}{-- $L_{ar}$}   & Twitter15 & 74.05  & 75.75  & 74.89 & -1.13 \\
			\multirow{3}{*}{}  & Twitter17 & 88.07 & 85.79 & 86.91 & -0.69   \\
			\multirow{3}{*}{}  & MRE & 82.48 & 80.93   & 81.70 & -1.53 \\
			\midrule
			\multirow{3}{*}{-- $L_{rr}$ \& $L_{ar}$}  & Twitter15    & 74.02 & 74.81  & 74.42 & -1.60\\
			\multirow{3}{*}{}  & Twitter17   &  87.15  & 85.86  & 86.50 & -1.10\\
			\multirow{3}{*}{}  & MRE  & 80.62 & 80.00 & 80.31   & -2.92 \\		
			\bottomrule
		\end{tabular}
	}
	\label{tab:test-ablation}
\end{table}
	
\subsection{Analysis for Modality-noise}

\begin{figure}[t]
		\centering
		\includegraphics[width=8.5cm]{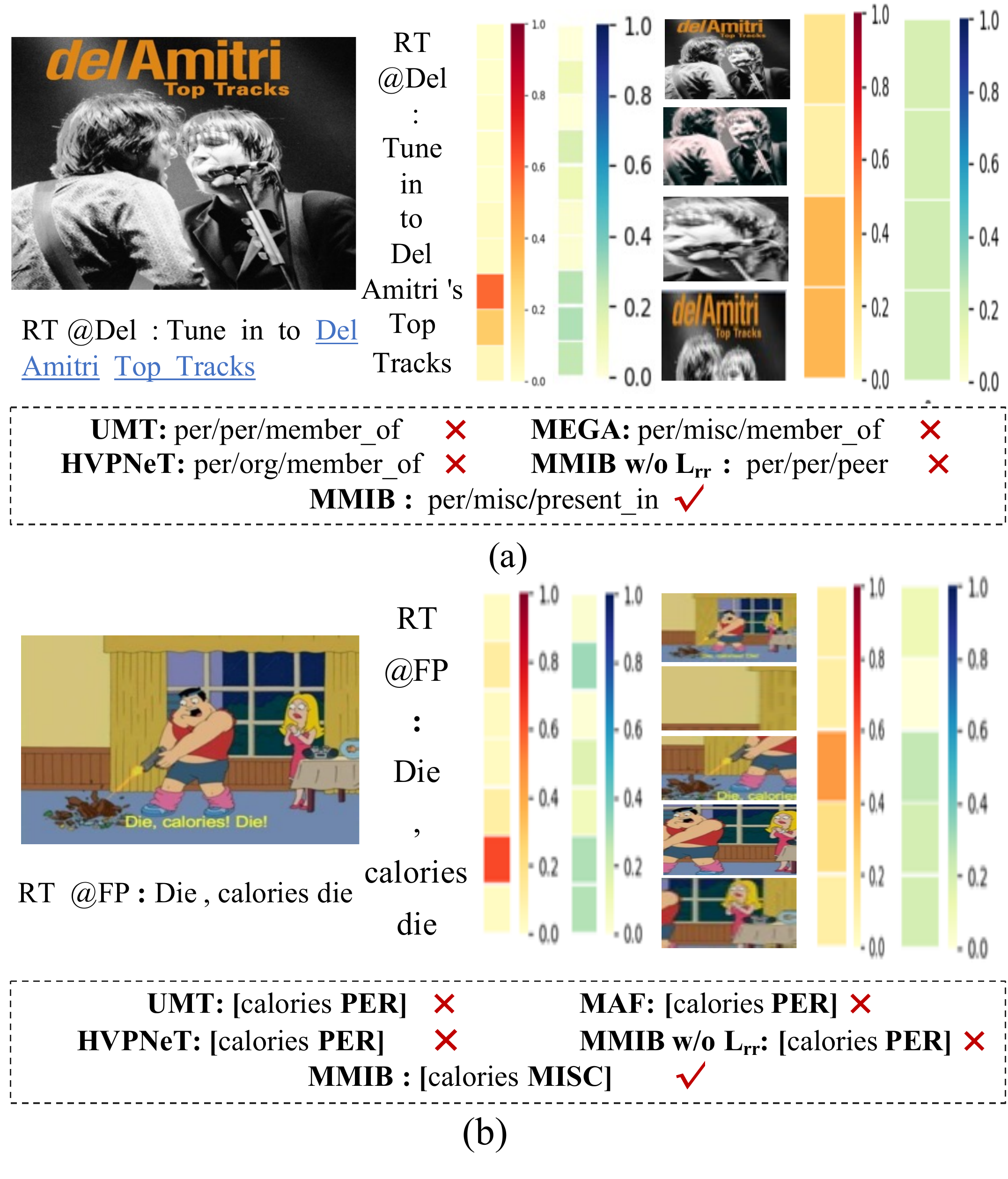}
		\caption{Case study for modality-noise.}
		\label{fig:case}
	\end{figure}

To confirm that MMIB derives text/image representations which  dismisses from redundant noises and maintain task-predictive, we present an visualization of two cases.
To quantify how much information the original representations $\mathbf{X}^T / \mathbf{X}^V$ attend to the learned representations $\mathbf{Z}^T / \mathbf{Z}^V$ , we define a contribution-score to measure it.
We illustrate the computation to contribution-score using the visual representations as an example.
Specifically, we first derive a matrix $\mathbf{A}^V \in \mathbb{R}^{(m+1) \times (m+1)}$ by dot-product between $\mathbf{X}^V / \mathbf{Z}^V$.
Then, we respectively sum each row of $\mathbf{A}^V$ obtaining $\hat{\mathbf{A}}^V \in \mathbb{R}^{m + 1}$, where $\hat{\mathbf{A}}^V_i$ is a scalar indicating how the $i_{th}$ image contributes to final visual representations $\mathbf{Z}^V$.
Following this process, we compute $\hat{\mathbf{A}}^V$, $\hat{\mathbf{A}}^T$ and visualize them.
The visualization to MMIB and MMIB w/o $L_{rr}$ are respectively deployed in the color map of red and green.
As Figure~\ref{fig:case}.(a) shows for MRE,  the task-irrelevant information (``RT @Del: Tune in to'') are weakened in MMIB.
As for the visual modality, in MMIB w/o $L_{rr}$, each image evenly attends to the final visual representations, which makes the attention mechanism in modality fusion wrongly incorporate the visual information of two singers, resulting in the prediction to $\mathtt{per/per/peer}$.
Meanwhile, such task-irrelevant visual information are dismissed in MMIB, which could facilitate the task prediction better.
For MNER in  Figure~\ref{fig:case}(b), we observe that the task-irrelevant information (i.e. the textual ``RT @FP'' and visual background in the image) attends little to the final image representations, where the redundant noises in each modality are dimissed.
With merit of $L_{rr}$, MMIB makes the correct prediction for these two cases.

Besides, we compare MMIB with some baselines  for performance analysis.
Specifically, UMT, one of vanilla multimodal baselines, predicts the relation as $\mathtt{peer}$ between persons in MNRE and assigns the entity type $\mathtt{PER}$ to ``calories'' in MNER.
Since UMT conducts the cross-modality interaction with all the grids in the image via the attention mechanism, the noisy grids could  disturb the attention assignments  and thus lead to the wrong predictions.
For MEGA, the noisy information could  distract the semantic alignment between the textual entities and visual objects.
MAF considers the relevance between the image and text globally, but the lack of fine-grained noise filter still produces the wrong prediction. 
%
Though HVPNet attempts to alleviate the visual noises by visual prefixes, it ignores the textual noisy semantics.
For example in Figure~\ref{fig:case}(b), the token ``die'' internally maintain semantics about a person, thus the visual prefix could be misled by the salient person objects.
Affected by such phenomenon, HVPNet fails to perform the correct predictions.

\begin{figure*}[t]
	\centering
	\subfloat[]{\includegraphics[width=0.32\textwidth,height=3.5cm]{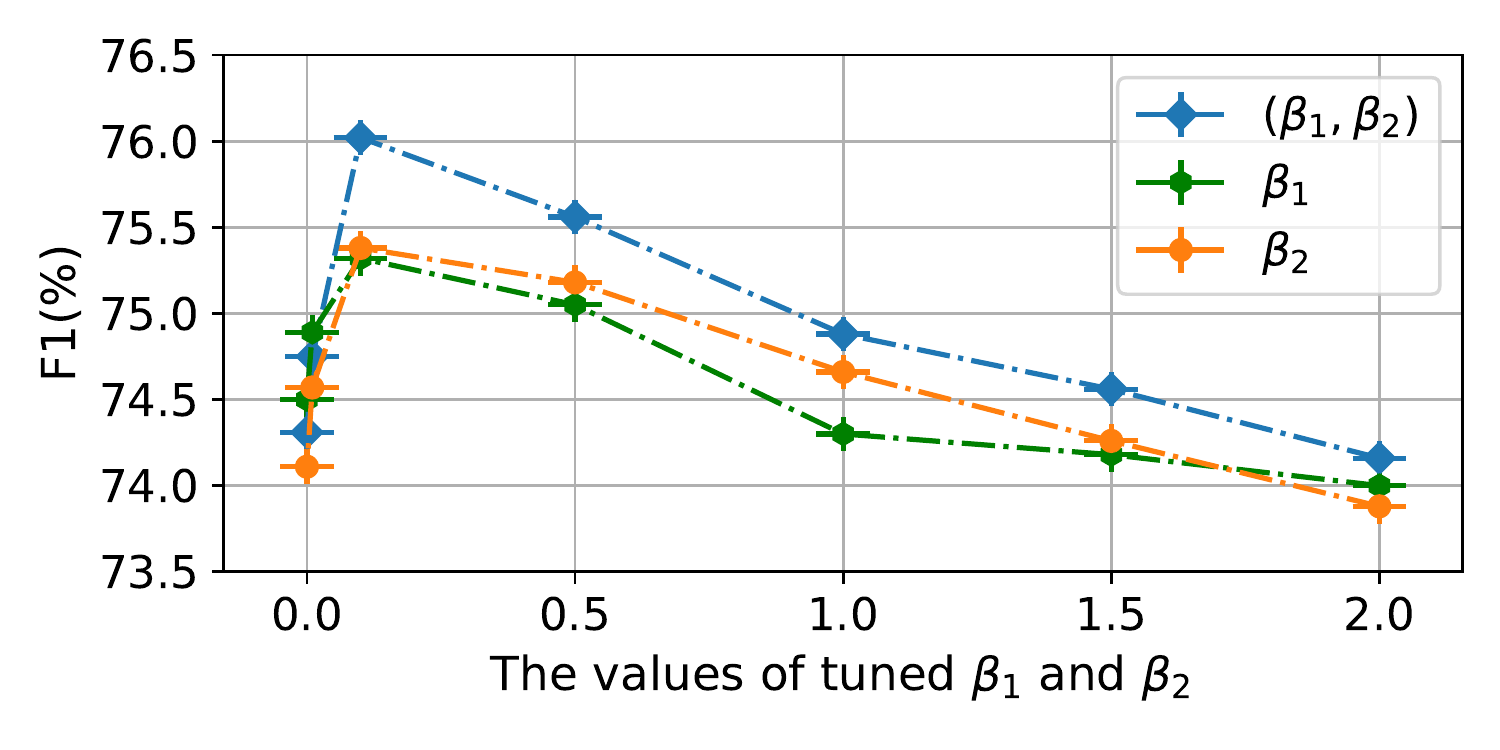}}
	\centering
	\subfloat[]{\includegraphics[width=0.32\textwidth,height=3.5cm]{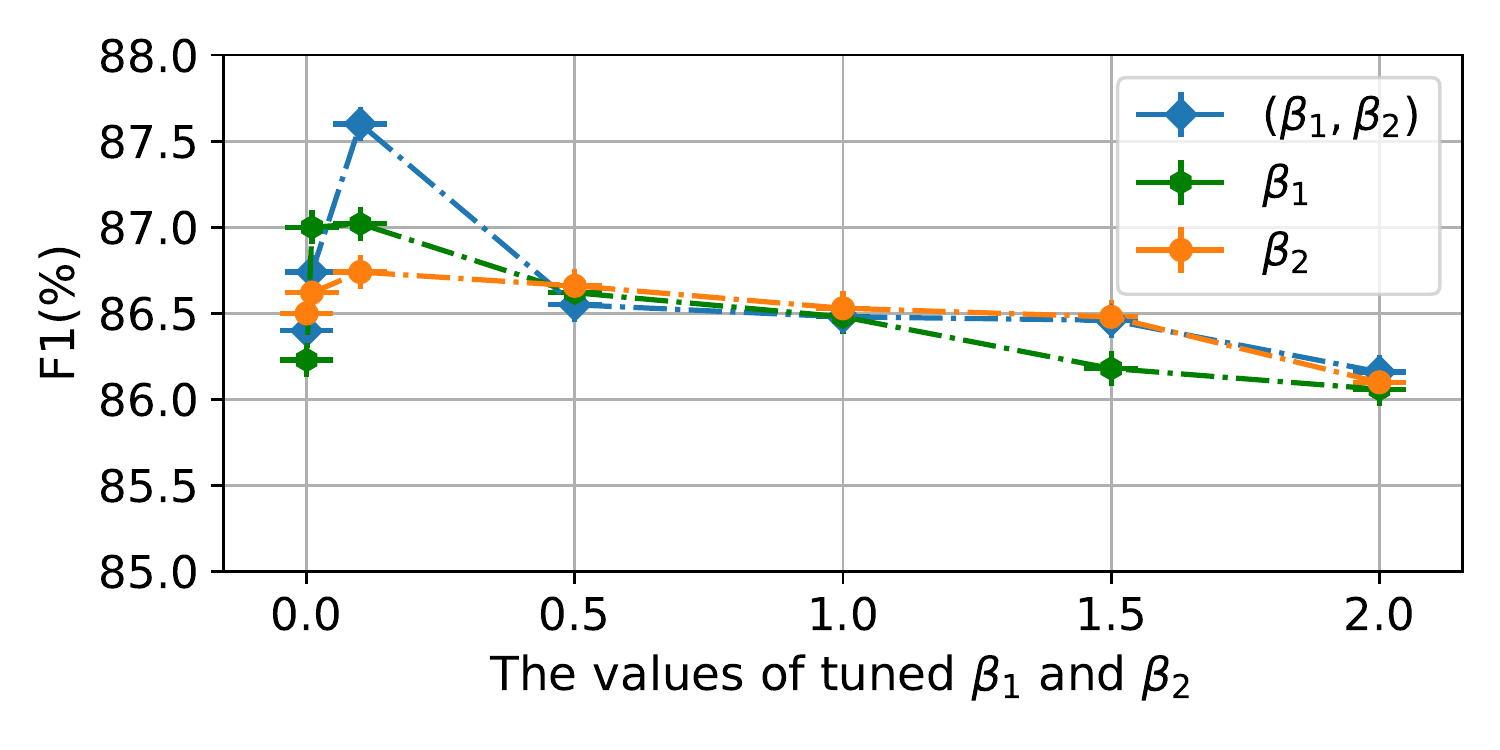}}
	\centering
	\subfloat[]{\includegraphics[width=0.32\textwidth,height=3.5cm]{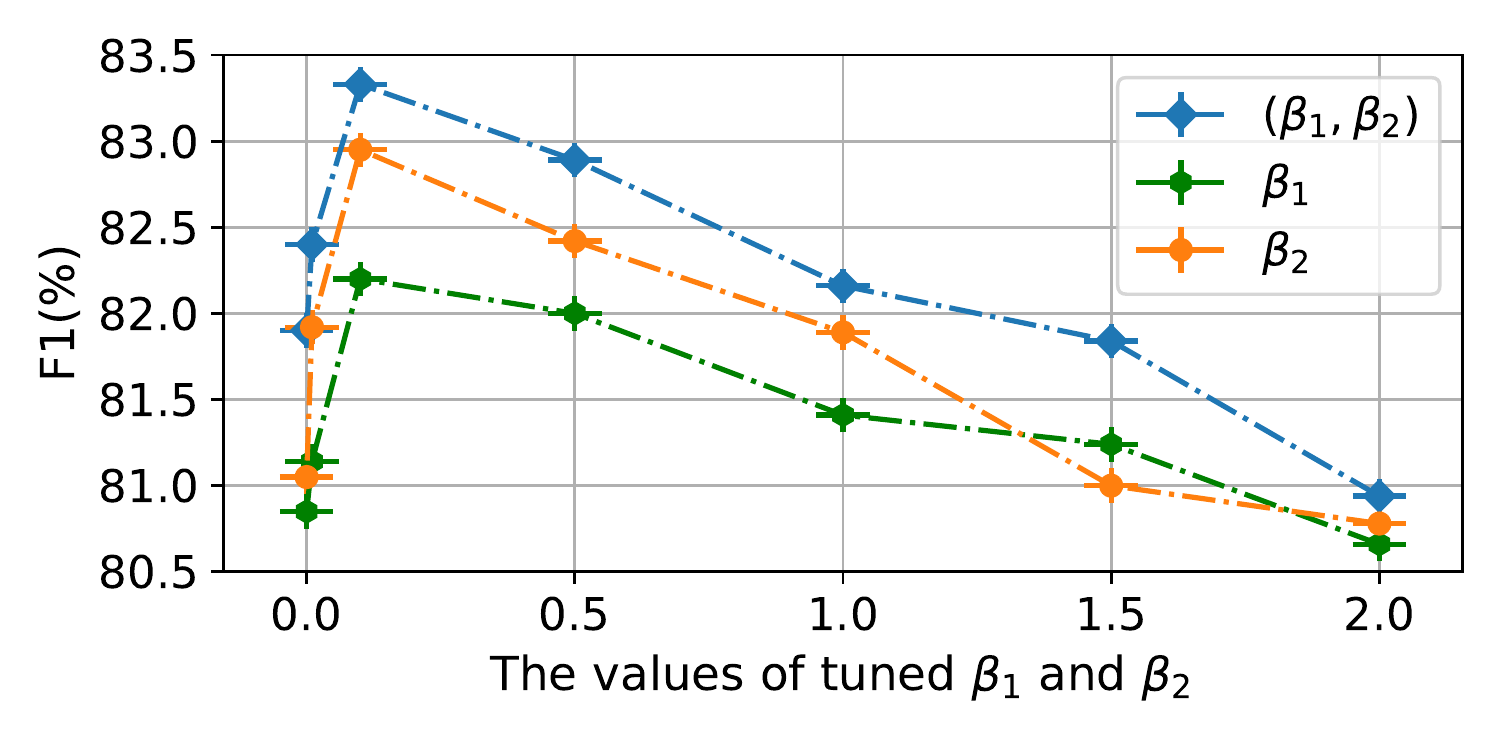}}
	\caption{The change F1 for our model with different values of $\beta_1$ and $\beta_2$ under different datasets: (a) The change of F1  on Twitter15 dataset; (b) The change of F1  on Twitter17; (c) The change of F1  on MRE dataset  }
	\label{fig:beta}
\end{figure*}

\subsection{Analysis to the Information Bottleneck}
To analyze how the information bottleneck principle works, we explore how the  Lagrangian multipliers, $\beta_1, \beta_2$ in $L_{rr}$, influence the final performances.
We employ $\{0,0, 0.01, 0.1, 0.5, 1.0, 1.5, 2.0\}$ as the candidate values of $\beta_1, \beta_2$. 
Based on these candidates, we conduct three series of experiments:
1) We fix $\beta_1=1$ and tune $\beta_2$ with the candidate values;
2) We fix $\beta_2=1$ and tune $\beta_1$ with the candidate values;
3) We tune $\beta_1, \beta_2$  simultaneously and keep their values as the same value.
The the fluctuation of $F1$ performances towards the experiments above are respectively presented with the orange, green and blue curves in Figure~\ref{fig:beta}.

Reading from Figure~\ref{fig:beta}, we could see that the best performances are achieved when both $\beta_1$ and $\beta_2$ are set as 0.1.
Specifically, we have observations and analysis as follows.
1) Tuning  $\beta_1, \beta_2$ simultaneously could generally achieve better performances than tuing them respectively, which demonstrates that these two IB-terms in $L_{rr}$ collaboratively work with each other to boost the final performances.
2) The performances degrade severely when $\beta_1$ or $\beta_2$ is set as 0.0.
Such a phenomenon manifests the importance of discarding noisy information from the original textual and visual information.
3) The worst performances are achieved when $\beta_1$ or $\beta_2$ is set as 2.0.
The reason might that the model compress the original textual and visual information too much, where little valid information is preserved to reconstruct the task-expected information.
4) The best performances are achieved with $\beta_1=0.1, \beta_2=0.1$, which reveals that  the noises in both modalities do not overweight the valid information.
Even so, it is important to discard the noises and control the trade-off between the noisy and valid information for predictions.

\begin{figure}[t]
		\centering
		\includegraphics[width=8cm,height=7.0cm]{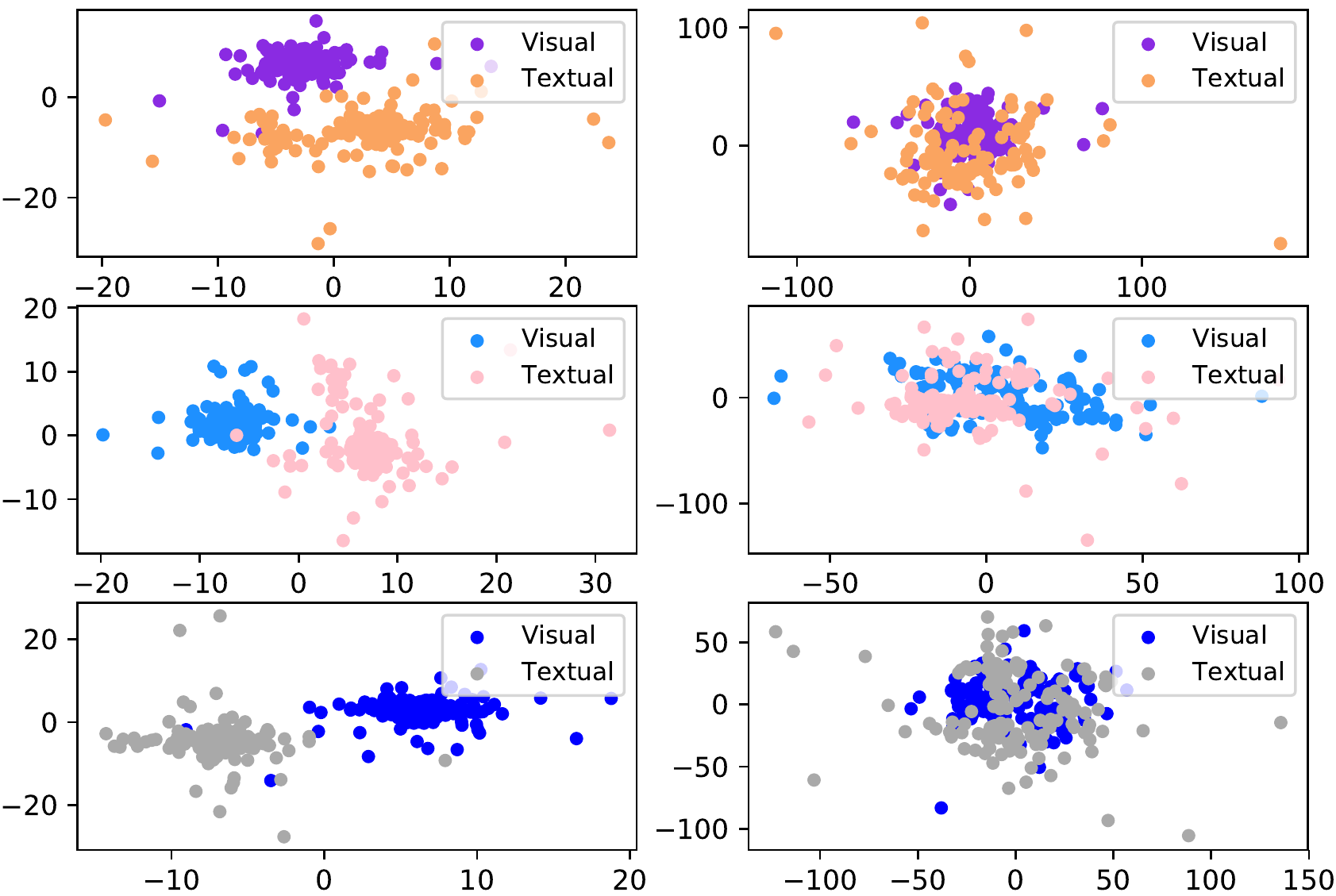}
		\caption{Representation visualization to modality-gap.}
		\label{fig:rep-tw17}
	\end{figure}

\subsection{Visualization for Modality-gap}
To confirm that MMIB produces the consistent text-image representations, we perform a text-image representation visualization. 
Specifically, we first randomly choose 150 pairs of entity-objects regarding PER, LOC and ORG types from the input text-image pairs, and gather their representations produced by trained MMIB with/without the alignment regularizer.
Then, t-SNE~\cite{DBLP:conf/nips/HintonR02} serves to visualize their representations, where it  is trained 500 iterations with perplexity of 10.
We deploy the results in Twitter17 dataset as an example and present the results in Figure~\ref{fig:rep-tw17}, where the three rows are respectively visualization to entity/objects of PER, LOC and ORG type. %
Without the alignment-regularizer, the entity/object representations respectively scatter in their own spaces as the left columns in Figure~\ref{fig:rep-tw17} shows.
On the contrary, the right column illustrates representation visualization obtained from the complete MMIB.
We could see that the alignment-regularizer could achieve the semantic agreements between textual and visual representations. 
This phenomenon manifests the effectiveness of the alignment-regularizer for consistent text-image representation learning.

\begin{figure}[t]
		\centering
		\includegraphics[width=8.6cm]{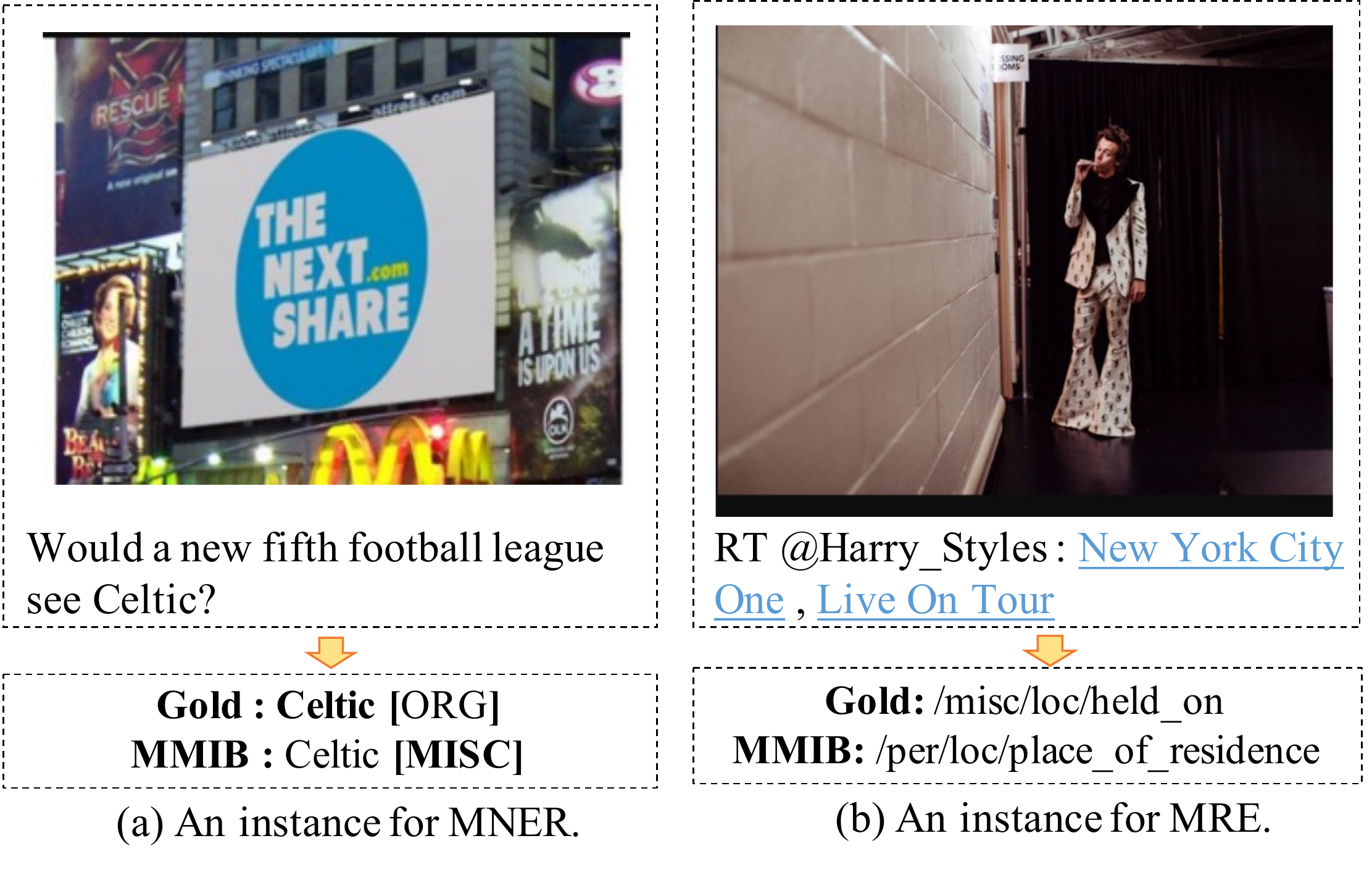}
		\caption{Error Analysis for the proposed MMIB model.}
		\label{fig:error}
\end{figure}

\subsection{Error Analysis}
To analyze the potential weakness of MMIB, we conduct an error analysis.
Specifically, we respectively  select 100 random mistakenly predicted instances  from three datasets, and observe that MMIB struggles  when the content of text and image are obviously irrelevant.
Figure~\ref{fig:error} shows some typical instances. 
For MNER in Figure~\ref{fig:error}.(a), the entity of interest ``Celtic'' refers to a foolball team of $\mathtt{ORG}$-type.
However, the salient object in the image is a billboard without obvious clues about an organization to a football team.
With such visual clues of the billboard, MMIB recognizes ``Celtic'' as $\mathtt{MISC}$-type. %
The similar situation also happens in  Figure~\ref{fig:error}.(b) for MRE.
Specifically, the sentence depicts the ``tour'' in ``New York City'', while the image poses objects about a person and a house.
Here, no obvious semantic correlations are presented between the content of text and image.
Hence, guided by such visual information, MMIB predicts $\mathtt{per/loc/place\_of\_residence}$ as the relation between given entities.

We attempt to infer the reasons for such phenomenon.
Despite that MMIB could discard the noisy information, some noisy information would be inevitably introduced when the content of text-image are obviously irrelevant.
We think that this points out an interesting direction and could inspire more future works to improve MRE and MNER.

\section{Related Work}
\subsection{Multimodal Entity and Relation Extraction}
Multimodal named entity recognition (\textbf{MNER}) and multimodal relation extraction (\textbf{MRE}) have raised great research interests due to the  increasing amount of the user-generated multimodal content.
Existing studies could be grouped into two lines.
The \textbf{first} line of studies directly incorporate the whole image as global visual clues to guide the textual task predictions.
The visual information are represented by encoding the image as either  one feature vector~\cite{DBLP:conf/naacl/MoonNC18,DBLP:conf/acl/JiZCLN18,DBLP:conf/aaai/0001FLH18} or multiple vectors corresponding to grids~\cite{DBLP:conf/wsdm/XuHSW22,DBLP:conf/acl/YuJYX20,DBLP:conf/aaai/0006W0SW21,DBLP:conf/aclnut/ChenANS21,DBLP:conf/icmcs/ZhengWFF021,DBLP:conf/coling/0023HDWSSX22}.
Correspondingly, attention-based mechanisms are designed to promote the text-image interaction, deriving expressive textual representations for the final tasks.
However, these methods struggle to build the fine-grained mapping between visual objects and named entities.
The \textbf{second} line of studies~\cite{DBLP:conf/mm/WuZCCL020,DBLP:journals/tmm/ZhengWWCL21} explore the fine-grained object-level visual information to boost the model performances. %
Specifically, these studies employ the pretrained object detection model  to extract the salient objects in the image, where the visual information are acquired from both the global image and local object image.
Apart from attention mechanisms~\cite{DBLP:conf/mm/WuZCCL020,DBLP:journals/tmm/ZhengWWCL21}, various strategies are proposed to sufficiently explore the text-image interaction.
For example, Zheng et al.~\cite{DBLP:conf/mm/ZhengFFCL021} and Zhang et al.~\cite{DBLP:conf/aaai/ZhangWLWZZ21} utilized multi-modal graph structure to capture the semantic correlation between texts and images.
Lu et al.~\cite{lu-etal-2022-flat} designed a unified lattice structure for cross-modal interaction.
Wang et al.~\cite{DBLP:journals/taslp/WangYLZL23} develop
scene graphs as a structured representation of the visual contents for semantic interaction.
To provide prior information about entity types and image regions, Jia et al.~\cite{DBLP:journals/corr/abs-2211-14739} and Jia et al.~\cite{DBLP:conf/mm/JiaSSPL00022} proposed machine reading comprehension (MRC) based methods, where the queries in MRC are elaborately designed to guide the cross-modality interaction for task predictions.

Though the studies above have achieved great success, modality-noise and modality-gap are two inadequately addressed issues. 
While most previous works ignore the modality-noise problem, Chen et al.~\cite{chen-etal-2022-good} attempt to prevent visual information from irrelevant objects via hierarchical visual features.
However, we focus on noises in both modalities and aim to derive representations which are not only noise-robust but also predictive enough for downstream tasks.
For the issue of modality-gap, though it could be bridged by converting the image or objects to their textual descriptions~\cite{DBLP:journals/tmm/ZhengWWCL21,DBLP:conf/mm/ZhengFFCL021,wang-etal-2022-ita}, such operation suffers from the error-propagation from external tools.
Different from prior works, we attempt to solve these two issues simultaneously via representation learning from the information-theoretic perspective, which is more generalizable and adaptive to other tasks.

\subsection{Information Bottleneck}
Information bottleneck (IB)~\cite{DBLP:conf/itw/TishbyZ15} is an important concept in information theory and has attached great research attention. 
IB is adapted into deep neural networks with variational inference, and thus is called variational information bottleneck (VIB)~\cite{DBLP:conf/iclr/AlemiFD017}.
VIB  could serve as a regularization technique for  representation learning and has shown its sparkles in various fields of computer vision~\cite{DBLP:conf/nips/ChenLGD18,DBLP:journals/tip/BarderaRBFS09,DBLP:conf/eccv/MotiianD16}, natural language processing~\cite{DBLP:conf/acl/ZhangZWZCH22,DBLP:conf/iclr/MahabadiBH21,DBLP:conf/acl/ZhouWCHH21,DBLP:conf/coling/ZhouZCHH22} and recommendation systems~\cite{DBLP:conf/icde/CaoSCLW22}.
Despite of the great achievements above, VIB has not been explored in MMNER and MRE yet, which is still unexplored and deserves the research attention.
In this paper, we adopt VIB to derive the task-expected and noise-robust representations for MNER and MRE,  and achieve great model  performances upon three public  benchmarks.

\section{Conclusion}
In this paper, we propose a novel approach, MMIB, for multimodal named entity recognition (MNER) and multimodal relation extraction (MRE).
MMIB aims to tackles the issues of modality-noise and modality-gap in these two tasks via representation learning with  information theory. 
Specifically, for the first issue, a refinement-regularizer, which explores the information-bottleneck principle, is devised to draw text-image representations which are diminished from the redundant noises and retain predictive for final tasks. 
For the second issue, an alignment-regularizer, which contains an MI-term, is proposed to derive the consistent cross-modality representations. %
Experiments on three benchmarks show that MMIB significantly outperforms state-of-the-art baselines.
In the future, we will adapt MMIB for other multimodal tasks for multimedia analysis.

\bibliographystyle{IEEEtran}
\bibliography{sample-base}

\end{document}